\documentclass[aps,prl,floats,amsmath,showpacs,twocolumn]{revtex4}
\usepackage{graphicx}

\begin{document}

\title{Amorphization induced by pressure: results for zeolites and
  general implications}

\author{Inmaculada Peral$^{1,2,3}$ and Jorge \'I\~niguez$^{1,3}$}

\affiliation{$^1$Institut de Ciencia de Materials de Barcelona
(ICMAB-CSIC), Campus UAB, 08193 Bellaterra, Spain\\ $^2$Departament de
F\'{\i}sica, Universitat Autonoma de Barcelona, 08193 Bellaterra,
Spain\\$^3$NIST Center for Neutron Research, National Institute of
Standards and Technology, Gaithersburg, Maryland 20899, USA}

\begin{abstract}
We report an {\sl ab initio} study of pressure-induced amorphization
(PIA) in zeolites, which are model systems for this phenomenon. We
confirm the occurrence of low-density amorphous phases like the one
reported by Greaves {\sl et al.} [Science {\bf 308}, 1299 (2005)],
which preserves the crystalline topology and might constitute a new
type of glass. The role of the zeolite composition regarding PIA is
explained. Our results support the correctness of existing models for
the basic PIA mechanim, but suggest that energetic, rather than
kinetic, factors determine the irreversibility of the transition.
\end{abstract}


\maketitle


Many minerals can be turned amorphous by mere application of pressure,
an intriguing phenomenon known as pressure induced amorphization
(PIA)~\cite{ric97}. The concrete realizations of PIA can be quite
diverse. Remarkably, upon release of the applied pressure some
materials recover their crystalline order, thus exhibiting reversible
amorphization, while others remain amorphous. Our understanding of PIA
is only partial. Simulations of systems like
$\alpha$-quartz~\cite{alpha-quartz}, silica~\cite{silica}, and
$\alpha$-berlinite~\cite{berlinite} suggest PIA is the result of a
first-order transition associated with very localized, weakly
interacting structural distortions that become unstable upon
compression. In such a transition, domain nucleation would overwhelm
growth and destroy the long-range order~\cite{coh02}. Cohen,
\'I\~niguez and Neaton~\cite{coh02} (CIN) further propose the PIA
transition will be reversible if the crystalline {\sl topology} is
preserved in the amorphous phase, i.e., if the amorphization does not
involve bond formation or breaking. While physically plausible, this
picture is yet to be confirmed.

The recent work of Greaves {\sl et al.} \cite{greaves} on the
nanoporous aluminosilicates known as zeolites has renewed the interest
in PIA. These authors have shown a zeolite may present two distinct
PIA phases: a low-pressure ($\approx$~2~GPa) low-density amorphous
phase (LDA), which they argue may constitute a new type of glass, and
a high-pressure ($\approx$~6~GPa) high-density amorphous phase
(HDA). Further, they claim the crystalline topology is preserved in
the LDA phase and lost in the HDA phase, which, according to the CIN
picture, implies amorphization will be reversible in the former case
and irreversible in the latter. These results, together with other
studies \cite{zeolites1,zeolites2} that, for example, show a striking
dependence of the PIA reversibility on the zeolite composition,
clearly point at these systems as ideal to test general PIA theories.

Here we present an {\sl ab initio} study of PIA in three
representative zeolites with different compositions. Our results (i)
confirm the occurrence of the above mentioned LDA and HDA phases, (ii)
show how the zeolite composition controls the nature of the PIA
transition, and (iii) essentially confirm, and perfect, the CIN
picture regarding PIA reversibility. While obtained for zeolites, our
results clearly pertain PIA phenomena at large.

\begin{figure}
\includegraphics[width=8cm]{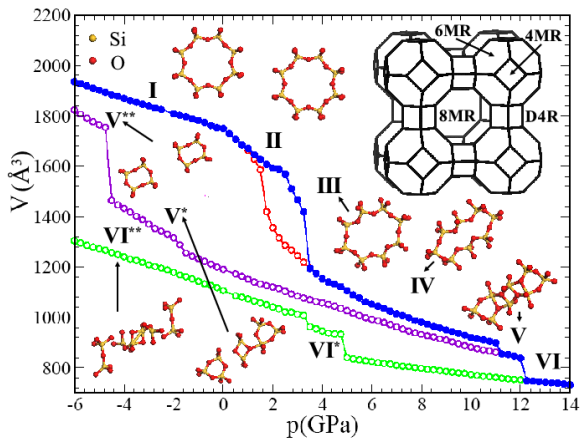}
\caption{Pressure dependence of the all-SiO$_2$ unit cell
  volume. Solid and open symbols refer, respectively, to compression
  and decompression. Also shown is the evolution of a representative
  eight-member ring. The top-right inset is a sketch of the
  LTA-framework structure as defined by the Si/Al atoms.}
\label{all-silica}
\end{figure}

We used the Generalized Gradient Approximation to Density Functional
Theory~\cite{dft} as implemented in the code SIESTA~\cite{calcs}. Note
that we wanted to study amorphization occurring in spite of neglegible
thermal activation, and thus focused on low temperature
simulations. We proceeded as follows: at each considered pressure, we
performed a short (100~fs) molecular dynamics simulation at 100~K,
starting with random velocities, and relaxed the resulting
structure. Zeolites have the general formula
$A^{n+}_{x/n}$Al$_x$Si$_{1-x}$O$_2$, where $A$ is a
charge-compensating cation (e.g. Li or Na). The Si and Al atoms are at
the center of corner-sharing O$_4$ tetrahedra. A typical zeolite
structure, with the so-called LTA framework, is sketched in the
top-right inset of Fig.~\ref{all-silica}; note the four-,
double-four-, six-, and eight-member {\sl rings} (denoted as 4MR's,
D4R's, etc.), also known as {\sl secondary building units}. We studied
three LTA zeolites with different percentages of Al and Na as the
charge-compensating cation: an Al-free ``all-SiO$_2$'' system, Na-ZK4
with a 1-to-5 Al-Si ratio, and Na-A where the ratio is
1-to-1. Following experimental information~\cite{zeolites3}, we
considered primitive cells containing 72, 76, and 168 atoms,
respectively, for all-SiO$_2$, Na-ZK4, and Na-A.

{\sl All-SiO$_2$ results}.-- The all-SiO$_2$ LTA zeolite exhibits most
of our key findings. Figure~\ref{all-silica} shows the pressure
dependence of the unit cell volume and the evolution of an 8MR that
captures the typical structural distortions. The slope changes and
volume discontinuities indicate a series of phase transitions. It is
sufficient for our purposes to describe the structure in terms of the
average values and standard deviations of the relevant angles
(Si--O--Si and O--Si--O) and distances (Si--O). For example, for the
reference LTA structure at 0~GPa we obtained $\bar{d}_{\rm
SiO}$=1.63$\pm$0.01~\AA, $\bar{\theta}_{\rm
OSiO}$=109$\pm$1$^{\circ}$, and $\bar{\theta}_{\rm
SiOSi}$=151$\pm$8$^{\circ}$.

Phases~I to III in Fig.~\ref{all-silica} are connected by continuous
transitions characterized by rigid rotations of the O$_4$
tetrahedra. Such Rigid Unit Modes (RUM's~\cite{ham94}) are mainly
reflected in $\bar{\theta}_{\rm SiOSi}$. For example, for phase~III at
2.75~GPa we obtained $\bar{d}_{\rm SiO}$=1.62$\pm$0.01~\AA,
$\bar{\theta}_{\rm OSiO}$=109$\pm$2$^{\circ}$, and $\bar{\theta}_{\rm
SiOSi}$=142$\pm$18$^{\circ}$. At 3.5~GPa phase~III transforms
discontinuously into a phase~IV of significantly smaller volume. This
transition does not involve any topological change, i.e. no bonds are
formed or broken. In fact, as in the previous cases, the structural
changes mostly affect $\bar{\theta}_{\rm SiOSi}$; at 7~GPa we
obtained: $\bar{d}_{\rm SiO}$=1.63$\pm$0.02~\AA, $\bar{\theta}_{\rm
OSiO}$=109$\pm$7$^{\circ}$, and $\bar{\theta}_{\rm
SiOSi}$=132$\pm$20$^{\circ}$.

At 11.25~GPa phase~IV transforms discontinuously into a phase~V, and
at 12.25~GPa there is another first-order transition to a
phase~VI. Phases~V and VI display collapsed rings of all types and new
Si--O bonds (see the 8MR depicted in Fig.~\ref{all-silica}), with the
corresponding loss of the LTA-framework topology. The occurrence of
SiO$_5$ and SiO$_6$ groups results in a wide dispersion of distances
and angles. For example, at 13~GPa we obtained $\bar{d}_{\rm
SiO}$=1.70$\pm$0.10~\AA, $\bar{\theta}_{\rm
OSiO}$=107$\pm$23$^{\circ}$, and $\bar{\theta}_{\rm
SiOSi}$=117$\pm$17$^{\circ}$. Note the increase in the average Si--O
distance, which reflects the existence of high-coordination defects.

Figure~\ref{all-silica} also shows our results regarding the
reversibility of the transtions. The transitions to phases~II and III
are reversible, with no hysteresis in the $V(p)$ curve. Phase~IV can
also transform back to the crystalline phase, but hysteresis occurs in
this case. Note that the presence or absence of hysteresis agrees with
the observed character, first- or sencond-order, of the
transition. Finally, upon decompression from phases V and VI, the
system undergoes a number of structural changes but does not find its
way back to the low-pressure stable phases. The transitions to phases
V and VI are thus irreversible.

Our simulations reveal the atomistic origin of this
irreversibility. The metastable phases V$^{**}$ and VI$^{**}$ in
Fig.~\ref{all-silica} present the right first-neighbor coordination
and can thus be viewed as formed by SiO$_4$ units. However, the ring
structure is not the LTA one. For example, in phase~V$^{**}$ the
original 8MR has transformed into two disconnected 4MR's, and phases
V$^*$ and VI$^{**}$ contain Si pairs sharing two oxygens (see
Fig.~\ref{all-silica}). Note that phases V$^{**}$ and VI$^{**}$ are
robustly metastable: while their excess energy with respect to phase~I
is relatively large (about 0.35~eV {\sl per} formula unit at 0~GPa), a
transition to the crystalline phase would require multiple Si--O bond
breakings within SiO$_4$ units, which is energetically very costly.

The origin of the defects affecting the ring topology can be easily
identified. Phases~V and VI exhibit collapsed rings in which bonds
form between Si and O atoms on opposite sides of the original
rings. The resulting SiO$_5$ and SiO$_6$ groups break upon
decompression, and the atoms recover their original
coordination. However, this defect breaking can happen in a variety of
ways from which {\sl only one} allows the ring to fully recover at low
pressures. In addition, this unique {\sl right way} involves the
largest volume expansion and, thus, is energetically favorable only at
relatively low pressures. For example, for phase~VI we obtained that
above 2~GPa the SiO$_5$ and SiO$_6$ groups find it energetically
favorable to break in ways that restore the original atomic
coordination but, at the same time, destroy the 8MR.

\begin{figure}
\includegraphics[width=5.6cm]{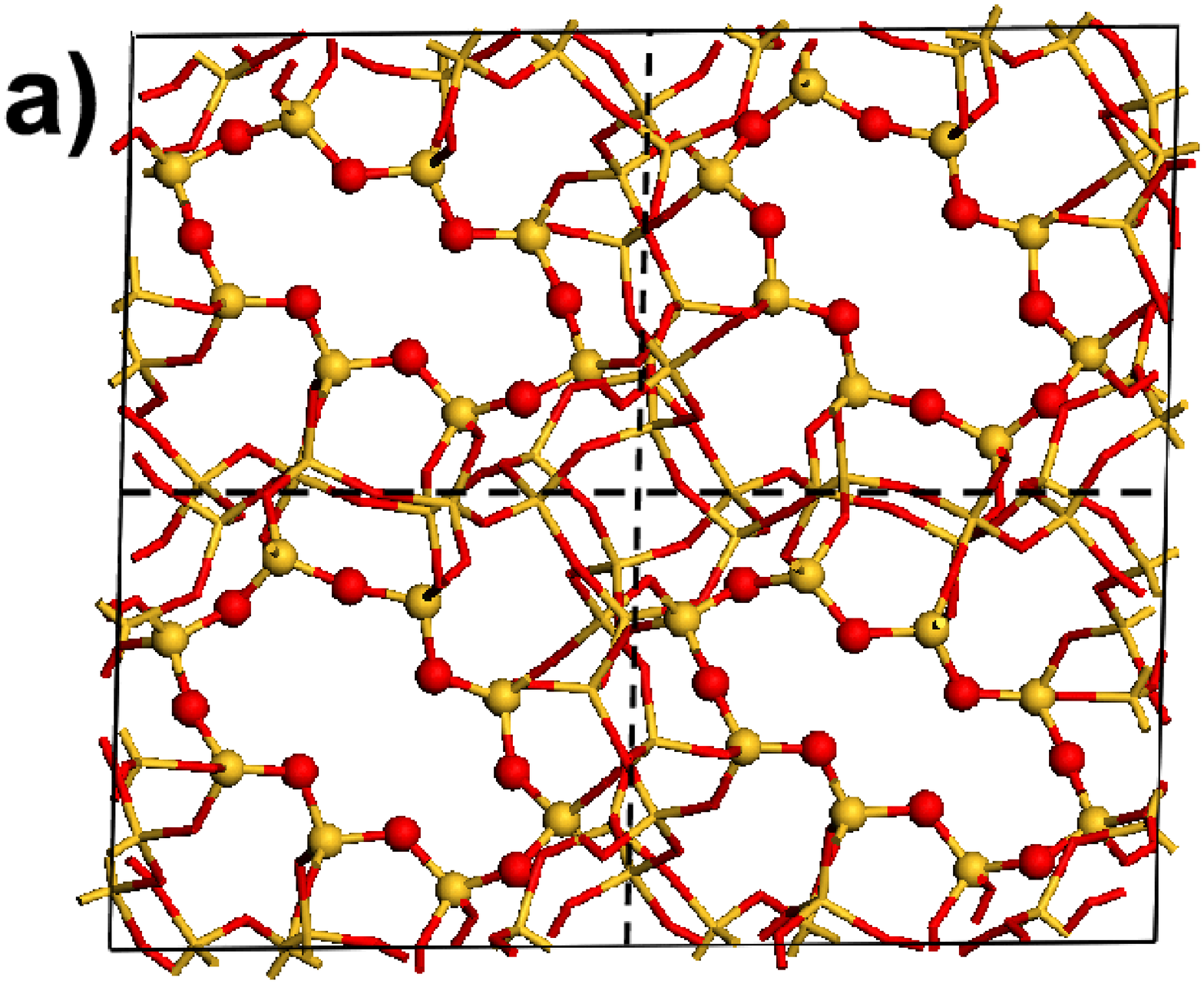}
\includegraphics[width=5.6cm]{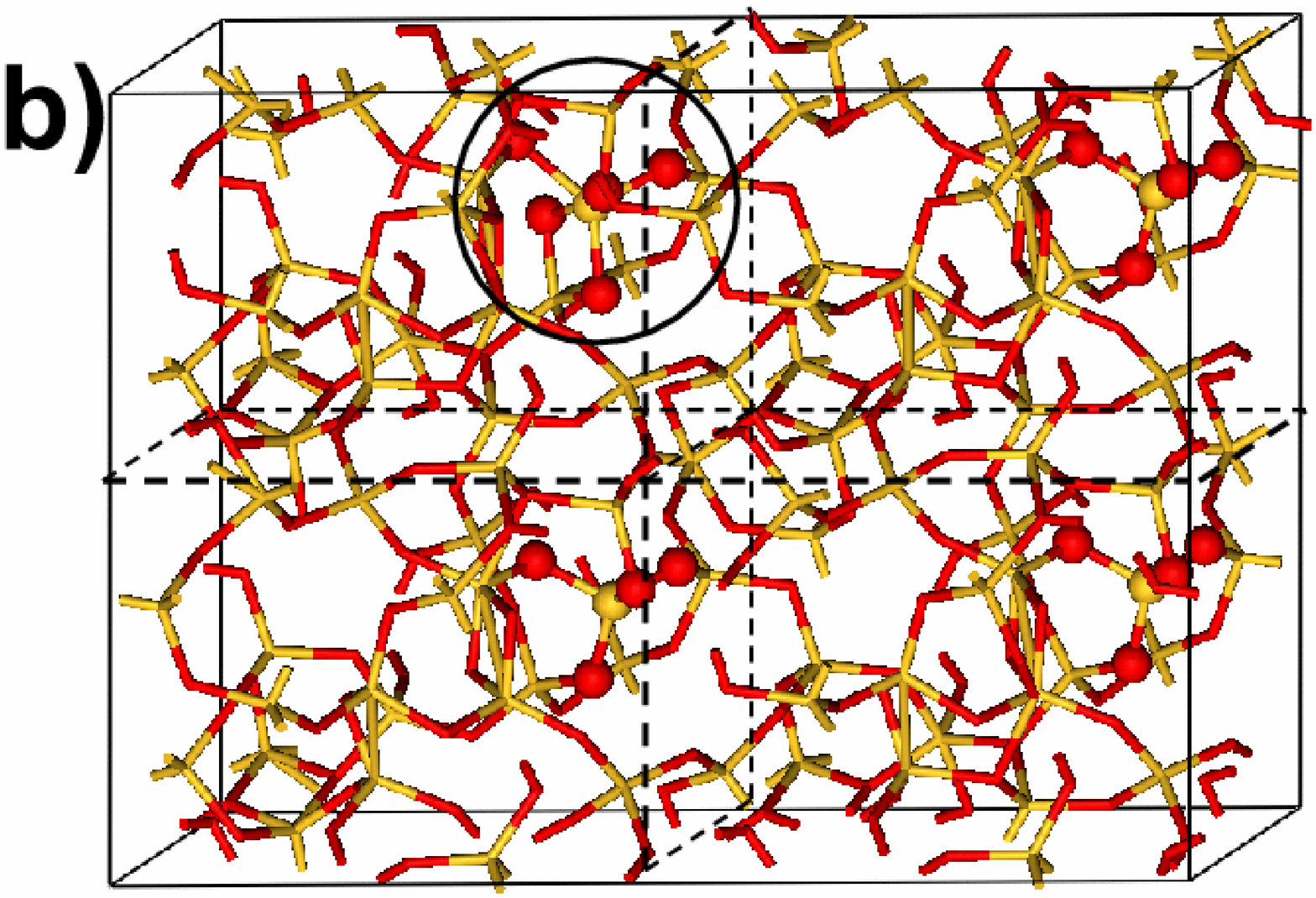}
\caption{All-SiO$_2$ 2$\times$2$\times$1 supercells (see text)
  resulting from transitions at 3.50~GPa (panel~a) and 11.25~GPa
  (panel~b). Dashed lines sketch the cell we start from. We highlight
  examples of atoms that were translationally related before the
  transition; a SiO$_5$ group is circled in panel~b.}
\label{amor}
\end{figure}

We also investigated if these transitions involve true
amorphization. Our simulated system is defined by the unit cell of the
LTA structure and, thus, cannot capture the translational symmetry
breaking characterizing amorphization. Hence, in a few selected cases
we considered larger supercells and simulated the loss of
translational symmetry directly. More precisely, we considered the
relaxed structures obtained at pressures neighboring transition points
(i.e. 0~GPa, 3.5~GPa, etc.), created the corresponding
2$\times$2$\times$1$\times$72-atom supercells, and increased the
pressure to observe how the transformation proceeds. Interestingly,
for the second-order transitions to phases~II and III, the
translations within the supercell are preserved, indicating that no
PIA occurs. On the other hand, as shown in Fig.~\ref{amor}, the
translational symmetry within the supercell is completely lost in the
transitions to phases~IV and V, both of which are first-order in
character. We can thus conclude that phases~IV and V are genuine
amorphous phases, the crystal topology being preserved in the former
and lost in the latter.


\begin{figure}
\includegraphics[width=8cm]{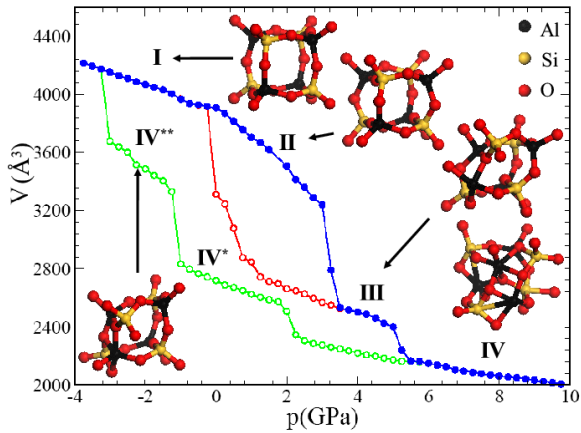}
\caption{Same as Fig.~\protect\ref{all-silica} for the Na-A zeolite.}
\label{naa}
\end{figure}

{\sl Na-ZK4 and Na-A results}.-- We consider first the case of Na-A,
which is more informative. Figure~\ref{naa} shows the pressure
dependence of the 168-atom primitive cell and the evolution of a D4R,
which captures the typical structural distortions. (Note that the Si
and Al atoms are intercalated in Na-A.) Phase~I is the reference
phase, which at 0~GPa is characterized by $\bar{d}_{\rm
SiO}$=1.65$\pm$0.01~\AA, $\bar{d}_{\rm AlO}$1.76$\pm$0.02~\AA,
$\bar{\theta}_{\rm OSiO}$=109$\pm$5$^{\circ}$, $\bar{\theta}_{\rm
OAlO}$=109$\pm$3$^{\circ}$, and $\bar{\theta}_{\rm
SiOAl}$=148$\pm$10$^{\circ}$. Phases~II and III are the product of two
transitions that are, respectively, continuous and discontinuous. Both
transformations are characterized by RUM's that involve rotations of
the O$_4$ tetrahedra and mainly affect the Si--O--Al angles. For
example, for phase~III at 5~GPa we obtained $\bar{d}_{\rm
SiO}$=1.66$\pm$0.05~\AA, $\bar{d}_{\rm AlO}$=1.82$\pm$0.09~\AA,
$\bar{\theta}_{\rm OSiO}$=109$\pm$10$^{\circ}$, $\bar{\theta}_{\rm
OAlO}$=108$\pm$21$^{\circ}$, and $\bar{\theta}_{\rm
SiOAl}$=118$\pm$18$^{\circ}$. Note that phase~III preserves the LTA
topology, clearly resembling what we found for phase~IV of
all-SiO$_2$.

At about 5.25~GPa the system undergoes a second discontinuous
transition to a phase~IV in which the D4R's collapse and the LTA
topology is lost. Comparison with all-SiO$_2$ clearly suggests the
presence of Al reduces the pressure at which the coordination defects
appear, which is consistent with the fact that Al is more likely than
Si to have a 5- or 6-fold oxygen coordination. Note also that only the
D4R's collapse in phase~IV. Interestingly, in Na-A the centers of all
the 6 and 8MR's are occupied by Na atoms, which suggests that the
cations are responsible for the preservation of the large rings. The
structure of phase~IV at 7~GPa is characterized by $\bar{d}_{\rm
SiO}$=1.67$\pm$0.06~\AA, $\bar{d}_{\rm AlO}$1.85$\pm$0.10~\AA,
$\bar{\theta}_{\rm OSiO}$=109$\pm$12$^{\circ}$, $\bar{\theta}_{\rm
OAlO}$=107$\pm$24$^{\circ}$, and $\bar{\theta}_{\rm
SiOAl}$=116$\pm$20$^{\circ}$. The relatively large standard deviations
reflect the structural disorder.

As shown in Fig.~\ref{naa}, we found that {\sl all} the transitions in
Na-A are reversible. Hysteresis does not occur in the case of the
second-order transition (to phase~II), but it does for the first-order
transitions (to phases~III and IV). Note that the reversibility from
phases~II and III, in which the ideal LTA topology is preserved, is
consistent with our results for all-SiO$_2$. However, the obtained
reversibility from phase~IV, in which the LTA topology is lost, is
clearly at odds with what we found in the all-SiO$_2$ case. Note also
that, unlike phases~V$^{**}$ and VI$^{**}$ of all-SiO$_2$,
phases~IV$^*$ and IV$^{**}$ of Na-A present coordination defects and
are not robustly metastable at low pressures.

Our results for Na-ZK4 (not shown here) can be summarized as follows:
(1) At low pressures there are continuous transitions dominated by
RUM's. (2) A first-order transition, at 4.25~GPa, causes coordination
defects, most of which involve neighboring Si-Al atom pairs that
approach to share two O atoms. All the Al atoms in the system give
raise to coordination defects, while only a small fraction of the Si
atoms do. (3) Upon further compression we find discontinuous
transitions to phases in which rings of all types collapse. (4) The
second-order transitions are reversible without hysteresis, and all
the first-order transitions are irreversible. Thus, as in Na-A, the
presence of Al in Na-ZK4 reduces the pressure at which the
coordination defects appear. On the other hand, at variance with
all-SiO$_2$ and Na-A, Na-ZK4 does not present any first-order
transition to a topology-preserving phase.


{\sl Discussion}.-- Our simulations render a wealth of conclusions
that pertain not only zeolites, but PIA phenomena at large. Maybe most
importantly, we confirm the existence of the LDA phases reported by
Greaves {\sl et al.}~\cite{greaves}. Two of the considered zeolites
(all-SiO$_2$ and Na-A) present an LDA phase, which supports the claim
that such phases may be quite common. Further, for all-SiO$_2$ the LDA
phase is predicted to be stable in a wide range of pressures. Our
results thus suggest that this zeolite, which has been recently
synthesized \cite{cor03} and is relatively simple, would be ideal for
detailed experimental and theoretical studies of the new type of glass
proposed in Ref.~\onlinecite{greaves}.

Our simulations confirm PIA is a first-order transition, thus
supporting the above mentioned nucleation-overwhelms-growth
mechanism~\cite{coh02}. In topology-breaking PIA transitions, the
structural distortions involve formation of new bonds and, as
Fig.~\ref{amor}b suggests, are rather localized and can freeze in
independently from the rest. In the PIA transitons that respect
topology, the localized distortions are essentially rigid rotations of
the O$_4$ tetrahedra. This is further confirmed by vibrational
calculations showing the {\sl whole} low-energy RUM-related band
softens under compression.

The Al atoms facilitate the formation of coordination defects. Such
defects are predicted to occur at about 5~GPa in Na-ZK4 and Na-A,
which is consistent with experimental data~\cite{greaves,zeolites1},
and only above 11~GPa in all-SiO$_2$. The Na cations, on the other
hand, impede the collapse of the rings at whose centers they are
located. As a result, there is no formation of new Si--O bonds between
atoms on opposite sides of Na-hosting rings. This seems consistent
with our finding that Na-A, which contains a large amount of Na,
presents an LDA phase while Na-ZK4 does not. Also, based on our
results, it seems reasonable to assume large cations will be more
effective in preventing rings from collapsing. That is in agreement
with reports that, for one particular zeolite, PIA is irreversible for
small cations (e.g. H) and reversible for larger ones (e.g. Li and
Na). The same rationale applies to the probable role of H$_2$O
molecules preventing PIA~\cite{zeolites2}.

Our results support the CIN picture~\cite{coh02} that
topology-preserving PIA transitions are reversible. However, at
variance with what is proposed in Ref.~\onlinecite{coh02}, they also
show that topology-breaking PIA transitions may be reversible. That is
the case of Na-A, where the topology breaking is caused by bond
formation between atoms that are close neighbors in the crystalline
phase. We find in such conditions the coordination defects can be {\sl
correctly undone} upon decompression, so that the crystalline
structure is recovered. We find the PIA transition to be irreversible
in cases in which the coordination defects involve atoms that are away
from each other in the crystalline phase (i.e. when the 6M and 8M
rings collapse). In such cases, the atoms recover their preferred
low-pressure coordination upon decompression, but the resulting ring
topology differs from the crystalline one.

Our simulations also refine the CIN picutre in what regards the
mechanism for irreversibility. According to Ref.~\onlinecite{coh02},
irreversibility occurs because the system is {\sl unable} to find the
{\sl transition path} to its most stable (crystalline) phase. To some
extend our results support such a kinetics-related explanation, as we
find that, for amorphous phases with collapsed rings, only a small
number of transition paths allow the system to recover both the ideal
coordination to first neighbors and the LTA ring structure. However,
our work also suggests a second and more important cause for the
irreversibility: We find that at high pressures (e.g. above 2~GPa) it
is energetically favorable to recover the ideal first-neighbor
coordination in ways that break the crystalline ring structure. Thus,
in addition to being more numerous, the {\sl wrong} transition paths
(i) are energetically preferred at moderately high pressures and (ii)
lead to phases that are robustly metastable at low pressures. This
{\sl thermodynamics-related} mechanism constitutes an alternative
cause, probably the main one in zeolites, for PIA irreversibility.

\acknowledgements

We thank J. Junquera for his help with the basis optimization and
R.L. Cappelletti for his comments on the manuscript. We thank the
finacial support of the Spanish Ministry of Science and Education
through the ``Ram\'on y Cajal'' program, MEC grants MAT2002-02808 (IP)
and BFM2003-03372-C03-01 (JI), the Catalan Government grant
SGR-2005-683 (JI), and FAME-NoE. We used the facilities of the CESGA
supercomputing center.

\end{document}